\documentclass[useAMS,usenatbib]{mn2e}

\usepackage{times}
\usepackage{graphicx}
\usepackage{url}
\usepackage{aas_macros}
\usepackage{amsmath}

\newcommand{\scinot}[2]{\ensuremath{#1 \times 10^{#2}}}

\newcommand{\paren}[1]{\left ( #1 \right )}
\newcommand{\bracket}[1]{\left [ #1 \right ]}

\newcommand{\parenfrac}[2]{\paren{\frac{#1}{#2}}}

\newcommand{\differd}[1]{\textrm{d}^{#1}}
\newcommand{\differ}[1]{\differd{}#1}

\newcommand{\deriv}[2]{\frac{\differ{#1}}{\differ{#2}}}

\begin{document}


\title[Gap-opening planets]{The Minimum Gap-opening Planet Mass in an Irradiated Circumstellar Accretion Disk}

\author[Edgar et al.]{
Richard G. Edgar,\thanks{rge21@pas.rochester.edu}
Alice C. Quillen,\thanks{aquillen@pas.rochester.edu}
\& Jaehong Park\thanks{jaehong@pas.rochester.edu} \\
Department of Physics and Astronomy, 
University of Rochester, Rochester, NY 14627}


\date{\today}

\pagerange{\pageref{firstpage}--\pageref{lastpage}}

\label{firstpage}


\maketitle

\begin{abstract}
We consider the minimum mass planet, as a function of radius, that is capable of opening a gap in an $\alpha$-accretion disc.
We estimate that a half Jupiter mass planet can open a gap in a disc with accretion rate $\dot M \la 10^{-8} M_\odot$/yr for viscosity parameter $\alpha =0.01$, and Solar mass and luminosity.
The minimum mass is approximately proportional to $\dot M^{0.48} \alpha^{0.8} M_*^{0.42} L_*^{-0.08}$.
This estimate can be used to rule out the presence of massive planets in gapless accretion discs.
We identify two radii at which an inwardly migrating planet may become able to open a gap and so slow its migration; 
the radius at which the heating from viscous dissipation is similar to that from stellar radiation in a flared disk, and the radius at which the disc becomes optically thin in a self-shadowed disc.
In the inner portions of the disc, we find that the minimum planet mass required to open a gap is only weakly dependent on radius.
If a migrating planet is unable to open a gap by the time it reaches either of the transition radii, then it is likely to be lost onto the star.
If a gap opening planet cuts off disc accretion  allowing the formation of a central hole or clearing  in the disc then we would estimate that the clearing radius would approximately be proportional to the stellar mass. 
\end{abstract}




\section{Introduction}

Recent observations have identified young (1--3 Myr old) stars 
that have inner clearings in the dust distribution as inferred from their 
IRS spectra (CoKuTau/4, \citealt{2005ApJ...621..461D}; 
GM Aur, TW Hya and DM Tau , \citealt{2002ApJ...568.1008C,2005ApJ...630L.185C}, 
and brown dwarf candidates L316 and L30003 in IC 348, \citealt{2006ApJ...643.1003M}). 
These disks have been dubbed ``transitional disks'' as
they represent the stage in which circumstellar disks are probably 
disappearing and in which massive planets are thought to be forming. 
In some cases there is still accretion on to the central star;
DM Tau and GM Aur at rates of 
$2\times 10^{-9}$ and $10^{-8}M_\odot$/yr, respectively. 
In other cases there is no
evidence for accretion, i.e., CoKuTau/4, even though its 
outer disk resembles that of other T-Tauri stars. 
Multi-band IRAC/MIPS photometric measurement from imaging surveys have
been used to identify dozens of other transition disk candidates 
\citep[e.g.][]{2006ApJ...638..897S}.
Candidates are then observed with IRS allowing accurate measurements 
of edge wall heights, and
temperatures and so radii \citep[e.g.][]{kim07}.

Two dominant approaches exist toward predicting and accounting for holes 
or clearings in young circumstellar disks. 
The first is planet formation followed by an opening of a gap and subsequent clearing
of an inner disk \citep[e.g.][]{2004ApJ...612L.137Q,2006ApJ...640.1110V,2007MNRAS.377.1324C}. 
Alternatively, photo-evaporation
\citep[e.g.][]{2001MNRAS.328..485C,2006MNRAS.369..229A}
may clear the disc from the inside out. 
A third possibility is that smaller dust particles are preferentially destroyed at a particular radius due to turbulence  \citep{2007ApJ...654L.159C}, within the context of a dead zone model. 
Photo-evaporation models can account for large disk clearings near luminous stars \citep{kim07} and non-accreting systems such as CoKuTau/4.
However they are not sufficiently sophisticated that they  can account for clearings in disks hosted by low luminosity brown dwarfs  \citep[UV flux is too low,][]{2006ApJ...643.1003M}),  disks with wide radial gaps and inner disks (such as GM Aur) or disks that are truncated in the dust distribution but continue to accrete (such as DM Tau).
Recently, \citet{2007arXiv0706.1241C} have suggested that stellar X-rays would be able to ionise the inner wall of a gap enough to trigger accretion via the magneto-rotational instability.
The disc would then clear from the inside out, at an accelerating rate.
Models involving a planet are likely to be more versatile but because of the complexity of planet/disk interactions  this scenario has been little explored.
In this paper we focus on the possibility that massive planets open gaps in disks and explore what mass planets are capable of opening gaps in different accretion disks.

Because a planet drives density waves into a disk
\citep{1979MNRAS.186..799L,1980ApJ...241..425G}
if the planet is sufficiently massive it can overcome
the effect of viscosity and open a gap in the disk
\citep{1993prpl.conf..749L,2000prpl.conf.1135W}.
\citet{2006Icar..181..587C} have generalized the gap opening criterion
to simultaneously take into account the mass of the planet, 
the scale height of the disk and the disk viscosity:
\begin{equation}
\frac{3}{4} \cdot \frac{H}{R_H}+\frac{50}{q R_{ey}} \leq 1 
\label{gc1}
\end{equation}
where $H$ is disk scale height, 
$R_H$ is the Hill radius, 
$R_{ey}$ is the effective turbulent Reynolds number and $q$ is the mass
ratio of planet mass $(M_p)$ to stellar mass $(M_*)$:%
\begin{displaymath} 
R_H    \equiv r_p \parenfrac{q}{3}^{1/3},\:
R_{ey} \equiv \frac{r_p^2 \Omega_p}{\nu},\: 
q      \equiv \frac{M_p}{M_*} 
\end{displaymath} 
Here $r_p$ is the planet's semi-major axis, $\nu$ is the disk viscosity and $\Omega_p$ is the angular rotation rate of an object in a circular orbit at $r_p$.
This criterion combines two earlier conditions; the viscous condition by
\citet{1999ApJ...514..344B}, and the tidal condition by
\citet{1993prpl.conf..749L}.

Gap opening has been studied extensively numerically 
with simulations
\cite[e.g.][]{1999ApJ...514..344B,2006Icar..181..587C,2006MNRAS.370..529D}.
However previous work has not predicted what mass planets are capable of opening gaps in circumstellar disks with different accretion rates or structure.
The gap opening criterion depends on the temperature profiles of the disk through  the dependence of the viscosity, $\nu$, on the sound speed, $c_s$, and vertical scale height, $H$.
Here we adopt an $\alpha$ form for the viscosity \citep{1973A&A....24..337S},
\begin{equation}
\nu = \alpha c_s H.
\label{eqn:nudef}
\end{equation}
Since accretion disc viscosity is poorly understood, $\nu$ and hence $R_{ey}$ are not precisely defined in this work;
this problem is common to all work on accretion discs.

The disk mid-plane temperature depends on the source of heating.
We consider viscous heating and heating from absorption of radiation from the central star.
Passive or non-accreting circumstellar disk models that  include the effect of radiation from the central star absorbed onto the disk have been studied by \citet{1997ApJ...490..368C,1999ApJ...526..411B,2004A&A...417..159D}.
Because of flaring, the irradiation of the disk star dominates the disk heating at large radii when both viscous heating and that from irradiation are considered \citep{1991ApJ...380..617C,2001ApJ...553..321D,2007ApJ...654..606G}.
Passive self-shadowed disks  can be considerably colder and thinner at large radii \citep{2004A&A...417..159D}.

In Section~\ref{sec:gapopening} we describe simple disk models and our procedure for estimating disk temperatures, scale heights.
From these, we can derive the minimum gap-opening planet mass  as a function of radius when heating is due to viscous dissipation and when heating is due to stellar irradiation.
Both cases are considered separately so we can be sufficiently flexible to discuss the case of a self-shadowed disc.
A disc that becomes self-shadowed will have significantly reduced heating due to stellar irradiation but heating from viscous dissipation would still be present.
Section~\ref{sec:results} presents our results, and a discussion follows in Section~\ref{sec:discuss}.

\section{Gap-Opening in accretion disks}
\label{sec:gapopening}

In this section
we apply the gap opening criterion \citep{2006Icar..181..587C}
to the $\alpha$-disk model of
\citet{1973A&A....24..337S} including the effect of stellar irradiation.

The accretion rate $\dot{M}$ of a steady thin disk can be calculated 
using mass and angular momentum conservation
\begin{equation} \dot{M} =3\pi\Sigma\nu. 
\label{ga12} 
\end{equation}
where $\Sigma$ is the disc surface mass density.
As the disk viscosity depends on the disk temperature
we must consider sources of heat to compute it.   
We use the $\alpha$ prescription 
to compute the viscosity (equation \ref{eqn:nudef})
but compute the sound speed and disk scale height
using the disk midplane temperature and
the relation for hydrostatic equilibrium, 
\begin{equation} 
    H=c_s/\Omega, 
\label{ga13}
\end{equation}
and the sound speed,
\begin{equation} 
    c_s = \sqrt{\frac{k_B T}{\mu m_H}}
\label{eqn:cs}
\end{equation}
where $k_B$ is the Boltzmann constant, $\mu$ is the mean molecular weight and 
$m_H$ is the mass of the hydrogen atom.
In our calculation, we take the mean molecular weight 
$\mu =2.4$ appropriate for interstellar gas.  

We first compute the structure of the disk taking
into account heat generated
from viscous dissipation. Then we consider the case of heat
generated from the radiation absorbed from starlight.


\subsection{Heat generated from viscous dissipation}

Dissipation due to viscosity gives the energy relation,
\begin{equation}
\frac{9}{8} \Sigma \Omega^2 \nu = \epsilon \sigma T_{\nu,s}^4,
\label{ga2} 
\end{equation}
where $\epsilon$ is the emissivity, 
$\sigma$ is the Stephan-Boltzmann constant. 
and $T_{\nu,s}$ is the temperature
at the disk surface.
A vertical average relates the surface temperature
to the midplane temperature, $T_{\nu,c}$
by $T_{\nu,c}^4 = {3 \over 8} \tau T_{\nu,s}^4$
where $\tau$ is the optical depth from the
surface to the midplane, $\tau=\kappa\Sigma/2$, 
and $\kappa$ is the opacity 
\citep[e.g., section 2 of][]{2001MNRAS.324..705A}.
We note that the relation between 
$T_{\nu,s}$ and $T_{\nu,c}$ in Equation~\ref{ga2} 
is only valid for optically thick disks.
In an optically thin region, we set $T_{\nu,c}$ to $T_{\nu,s}$.

We adopt a convenient analytic form of the emissivity
and opacity laws
that is based on the assumption that dust grains govern 
the opacity and emissivity, $\epsilon = (T/T_\odot)^b$ and
$\kappa = \kappa_V (T/T_{\odot})^b$ 
where $\kappa_V$=1~cm$^2$g$^{-1}$, $T_\odot$ 
is the solar effective surface temperature and $b=1$
\citep{1997ApJ...490..368C,2007ApJ...654..606G}.
This form is expected as the dust temperature over 
much of the disk corresponds to a peak wavelength (for a black
body spectrum)
that is larger than the diameter of the dust particles 
and the diameter is of order the peak wavelength of a Solar temperature
black body 
\citep[e.g.][]{1992ApJ...385..670B,1997ApJ...490..368C}.
Using Equations~\ref{eqn:nudef}), \ref{ga12}---\ref{ga2} 
and the opacity law, 
we solve for the midplane temperature finding
\begin{equation}
T_{\nu,c} \approx 
\left\{
\begin{array}{ll}
       \bracket{ \frac{3}{128 \pi^2} \frac{\mu m_H}{k_B}
           \frac{\kappa_V}{\sigma\alpha } \Omega^3 \dot{M}^2
      }^{\frac{1}{5}} &  {\rm for} ~~ \tau \gg 1  \\
      \bracket{ \frac{3}{8 \pi}  \frac{\dot{M} \Omega^2 T_\odot}{\sigma}
      }^{\frac{1}{5}} &  {\rm for} ~~ \tau \ll 1  \\
\end{array}
\right.
\label{ga3}
\end{equation}
The opacity drops to the low opacity regime
at an approximate dividing line of $\tau=8/3$ at a radius
\begin{equation}
\begin{split}
r_{\tau} &\sim
     \bracket{
                   \frac{\kappa_V}{16 \pi}
	           \frac{\mu m_H}{k_B T_\odot} 
                   \frac{\dot{M}}{\alpha}
             }^{\frac{2}{3}}
            \paren{GM_*}^{\frac{1}{3}} \\
      & \sim
          1.2 \, \textrm{AU} \,
          \parenfrac{\dot{M}}{10^{-8} M_\odot/{\rm yr}}^{\frac{2}{3}}
          \parenfrac{\alpha}{0.01}^{-\frac{2}{3}}
          \parenfrac{M_*}{M_\odot}^{\frac{1}{3}}
\end{split}
\label{eqn:rtau1}
\end{equation}

For the $\alpha$ disk model, using the above equations,
we solve for all disk variables (temperature, density, viscosity,
scale height) as functions of parameters
$\alpha$, $\dot{M}$, and $M_*$ and as a function of radius, $r$.

\subsection{Heating due to irradiation from the star}

The disk can be heated not only by disk viscosity 
but also by the stellar radiation.
In this section, we consider the latter case only.
If we regard the central star as a point source, then 
the surface temperature depends on the slope of
the disk,
\begin{equation} 
\frac{L_*}{4\pi r^2} (1-\beta) \frac{H}{r} \paren{\deriv{\ln H}{\ln r}-1}
= \epsilon \sigma T^4_{i,s}, 
\label{irr1}
\end{equation}
\citep{2002apa..book.....F}, 
where $L_*$ is the stellar luminosity, 
$\beta$ is the albedo 
and $T_{i,s}$ is disk surface temperature by irradiation.

Here, we assume that the disk temperature
does not strongly depend on height, though a more
detailed model would more carefully compute
the disk structure
\citep[e.g.][]{1998ApJ...500..411D,2007ApJ...654..606G}.
Indeed, if the disc is very optically thick, then the midplane temperature will always be dominated by viscous dissipation \citep{1999MNRAS.303..139D}.
Ignoring the vertical temperature gradient, we estimate
the disk midplane temperature $T_{i,c}\simeq T_{i,s}$ using Equations~\ref{ga12}---\ref{ga2} and~\ref{irr1}:
\begin{equation} 
   T_{i,c}  \sim
    \parenfrac{A T_\odot}{\sigma}^{\frac{2}{9}}
    \parenfrac{k_B}{\mu m_H}^{\frac{1}{9}}
    \paren{G M_*}^{-\frac{1}{9}}
    r^{-\frac{1}{3}}
    \label{irr2}
\end{equation}
where the coefficient
\begin{displaymath} 
   A=\frac{L_*}{4\pi}(1-\beta)
    \paren{\deriv{\ln H}{\ln r}-1}
\end{displaymath}

We restrict our solution for the disk scale height to a self-consistent
power law form with ${d\ln H/d\ln r}$ equal to a constant. 
We find that the disk scale height
\begin{equation}
   H \sim \parenfrac{L_* (1-\beta) T_\odot}{12 \pi\sigma}^{\frac{1}{9}} 
          \parenfrac{k_B}{\mu m_H}^{\frac{5}{9}} 
          (G M_*)^{-\frac{5}{9}} 
          r^{\frac{4}{3}} 
\label{irr3},
\end{equation}
and that ${d\ln H / d\ln r}={4/3}$. 
If we consider an albedo in the range $0<\beta<0.5$ \citep{2002ApJ...567.1183W},  then Equations~\ref{irr2} and~\ref{irr3} imply that the disk structure is not strongly sensitive to the albedo. 
In our subsequent estimates, we have adopted an albedo of $\beta=0$.

Interior to a particular radius, the heat from viscous heating dominates that from stellar irradiation.
Setting the heating rate due to accretion equal to that due to irradiation   
(setting the left hand side of Equation~\ref{ga2} to the left hand side of Equation~\ref{irr2} and solving for radius) we estimate a transition radius
\begin{equation}
\begin{split}
r_{tr} & \sim
           \parenfrac{9}{2}^{\frac{3}{4}}
           \parenfrac{12 \pi \sigma}{T_{\odot}}^{\frac{1}{12}}
           \paren{ L_* (1-\beta) }^{-\frac{5}{6}}
           \times \\
       & 
           \parenfrac{ \mu m_H}{k_B}^{\frac{5}{12}}
           \dot{M}^{\frac{3}{4}} 
           \paren{GM_*}^{\frac{7}{6}} \\
       & \approx
          0.3 {\rm AU} 
          \parenfrac{\dot{M}}{10^{-8} M_\odot/{\rm yr}}^{\frac{3}{4}}
          \parenfrac{M_*}{M_\odot}^{\frac{7}{6}}
          \parenfrac{L_*}{L_\odot}^{-\frac{5}{6}}
\end{split}
\label{eqn:rtr} 
\end{equation}
For $r<r_{tr}$ we expect that viscous heating would dominate.

\section{Results}
\label{sec:results}

\begin{figure}
\includegraphics[angle=0,width=3.0in]{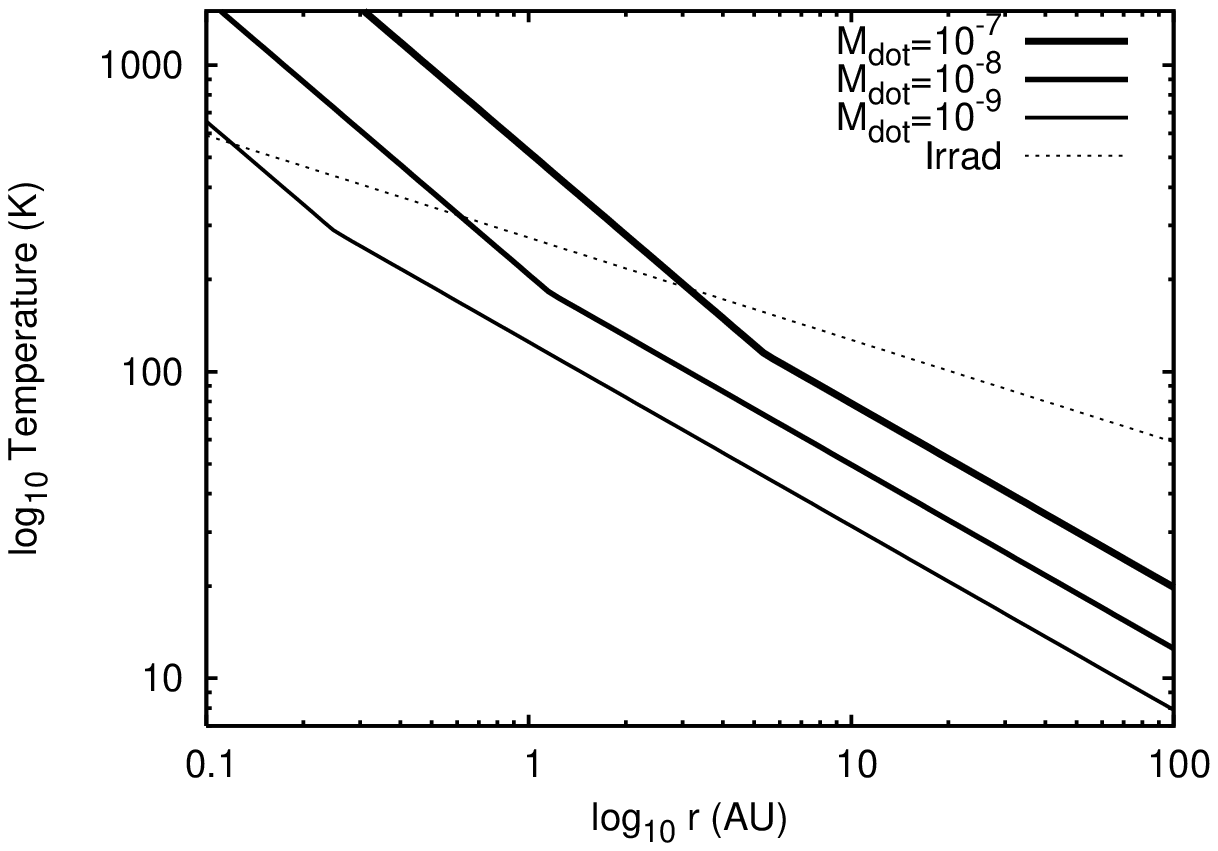}
\includegraphics[angle=0,width=3.0in]{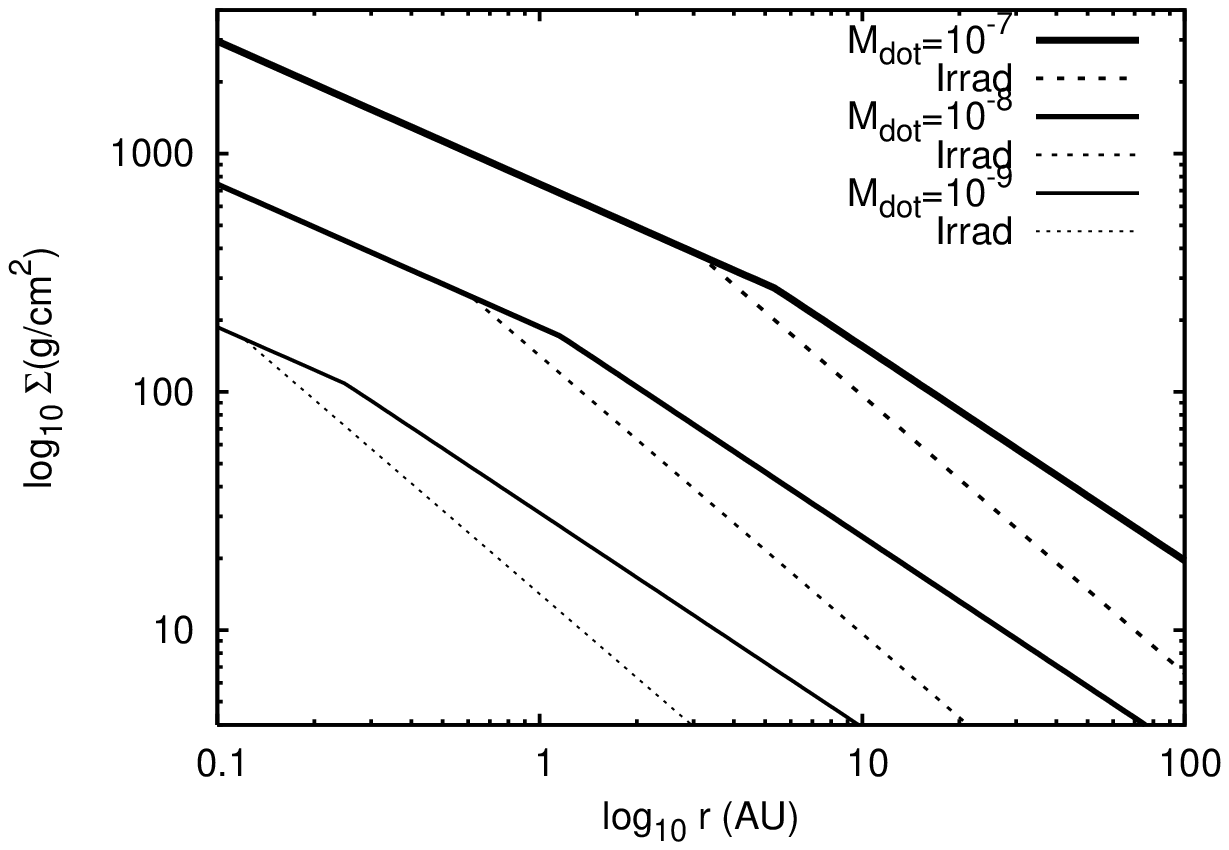}
\includegraphics[angle=0,width=3.0in]{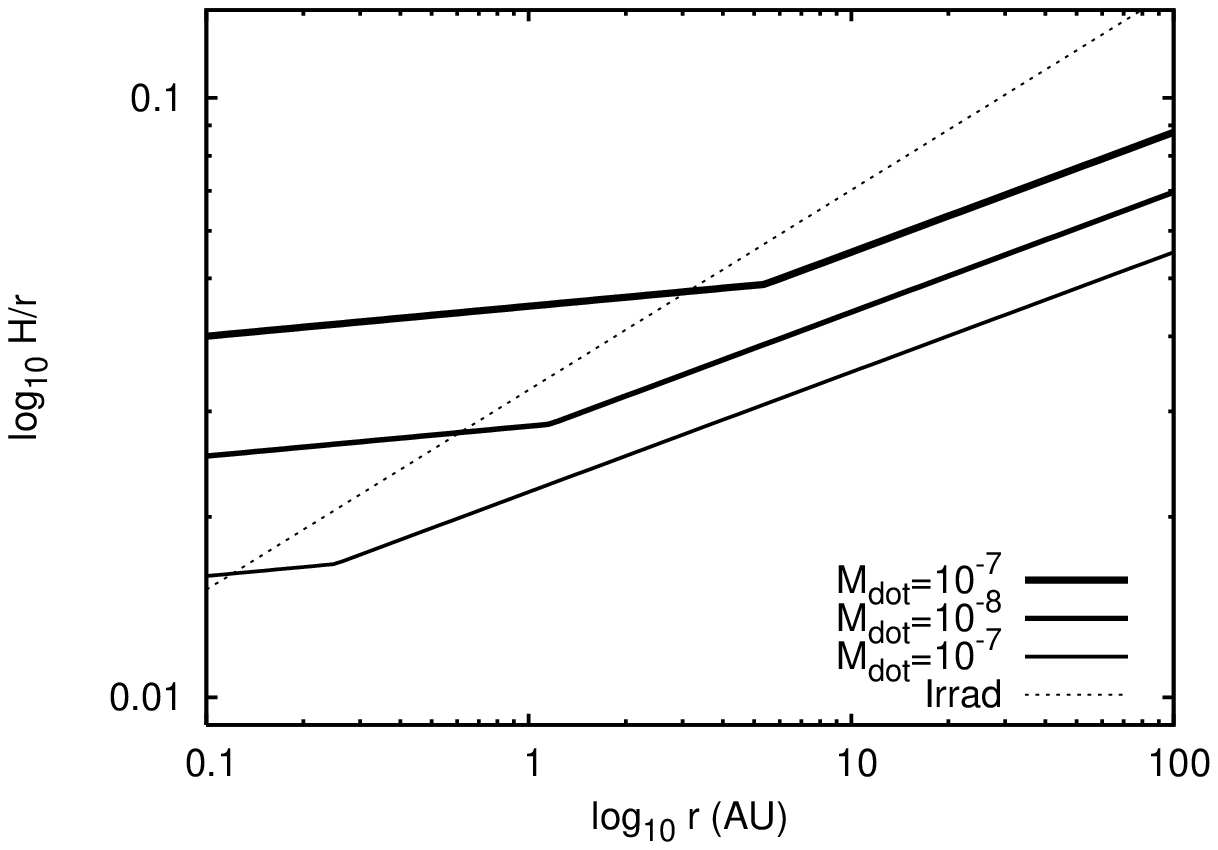}
\caption{\label{fig:diskprofiles} 
a) Disk radius vs.  midplane temperature 
for disk mass accretion rates
$\dot{M}=10^{-9}, 10^{-8}$ and $10^{-7}$ $M_{\odot}$/yr for $\alpha=0.01$ 
and stellar mass $M_* =1 M_{\odot}$.
The solid lines are for the case of viscous heating only and 
the dotted lines are for the case of irradiated heating only
and $L_* = L_\odot$.
A comparison between the dotted and solid lines shows
that heating due to stellar radiation dominates at large radius.
A change in slope in the solid lines occurs at the radius where
the disc becomes optically thin ($r_{\tau}$).
b) Similar to a) except the surface density profile is shown.
c) Similar to a) except the aspect ratio $H/r$ is shown.
}
\end{figure}

In Figure~\ref{fig:diskprofiles} we show disk variables as a function of radius 
for different accretion rates $\dot M = 10^{-7},10^{-8}$, and $10^{-9} M_\odot$/yr but fixed $\alpha = 0.01$ and  stellar mass, $M_*=1 M_\odot$. 
Figure~\ref{fig:diskprofiles}a shows the disk mid-plane temperature, 
Figure~\ref{fig:diskprofiles}b shows the surface density profile and
Figure~\ref{fig:diskprofiles}c the vertical scale height for these disks, and as a function of radius in each case.
In our Figures, we show with solid lines the variables calculated for heating primarily from viscous dissipation while variables primarily due to the stellar radiation are shown with dotted lines.
A comparison between the dotted and solid lines shows that heating due to stellar radiation dominates at large radius whereas heating due to viscous dissipation dominates at small radius.
In Figure~\ref{fig:diskprofiles}a-c the change in slope in the solid lines  occurs where the disc becomes optically thin at a radius approximately given by $r_\tau$ (Equation~\ref{eqn:rtau1}).

We solve for the minimum gap opening planet mass ratio using Equation~\ref{gc1}.
With a variable substitution of $y=(q_0/q)^{\frac{1}{3}}$
equation \ref{gc1} can be written as a cubic equation
\begin{displaymath}
B y + y^3 = 1
\end{displaymath}
where we have defined
\begin{equation*}
q_0 \equiv \frac{50}{R_{ey}}, \,
B =  \frac{3}{4} \cdot \parenfrac{3 h}{50 \alpha r}^{\frac{1}{3}}
\end{equation*}
The cubic equation has only one real root
\footnote{See \url{http://mathworld.wolfram.com/CubicFormula.html}}
\begin{displaymath}
y= \frac{B}{3 u} - u
\end{displaymath}
with 
\begin{displaymath}
u^3 = -\frac{1}{2} + \sqrt{\frac{1}{4} + \frac{B^3}{27}}.
\end{displaymath}
For the entire parameter space covered here we find that $B< 1$ and the minimum gap opening planet mass $q_{min} \approx q_0$.
Although the prediction of \citeauthor{2006Icar..181..587C} (Equation~\ref{gc1} in this paper) is more general, we find that for most disc models the viscous condition of previous work suffices to estimate the minimum gap opening planet mass.
We would not expect this to be the case in discs containing `dead zones' \citep{1996ApJ...457..355G}, where the viscosity is low, but the scale height remains large.

\begin{figure}
\includegraphics[angle=0,width=3.0in]{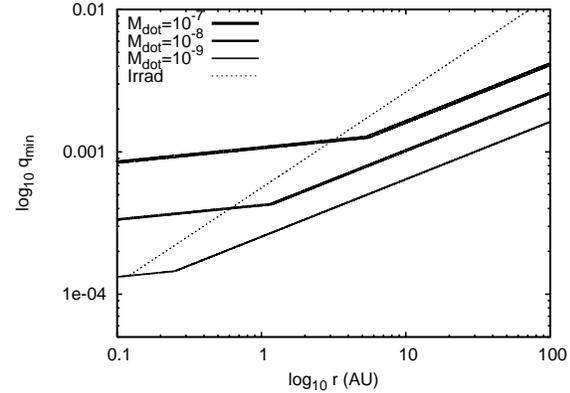}
\caption{\label{fig:minimumplanetmdot} 
The minimum gap-opening planet mass ratio as a function
of radius (computed using Equation~\ref{gc1}) for the
disks shown in Figure~\ref{fig:diskprofiles} with accretion
rates $\dot M = 10^{-7}, 10^{-8}$, and $10^{-9}M_\odot$/yr.
The transition radius is evident where the dotted line
intersects the solid lines.  This radius is where
midplane temperature from viscous dissipation is similar to that
from stellar radiation.  Inside this radius the minimum
gap-opening planet mass is not strongly sensitive to radius.
Outside this radius a larger planet mass
is required to open a gap, as long as the disk is
sufficiently flared to be heated by starlight.
Lower mass planets can open a gap at the larger transition radius
(slope change of the solid lines) in self-shadowed disks.  
}
\end{figure}

Figure~\ref{fig:minimumplanetmdot} shows the minimum gap-opening planet mass ratio for these disk models as a function of radius.
In the inner regions, where viscous heating dominates, the minimum $q$ required to open a gap is not strongly dependent on radius.
However, at larger radii, the minimum gap opening $q$ begins to rise more steeply with increasing radius.
This happens for both self-shadowed and irradiated discs.
If the disc is self-shadowed, then we would expect the minimum planet mass for gap opening always to lie along the solid (viscous heating) line.
As noted in Section~\ref{sec:gapopening}, \citet{1999MNRAS.303..139D} showed that the midplane temperature of a very optically thick irradiated disc would be dominated by viscous dissipation.
In this case, the minimum planet mass required to open a gap would lie somewhere between the solid (viscous) and dotted (irradiated) lines.
However, unless the disc is self-shadowed, the curve must rejoin the dotted line around $r_{\tau}$, since then the radiation will be able to penetrate to the midplane.

Figure~\ref{fig:minimumplanetmdot} may be compared to figure~7 of \cite{2004ApJ...606..520M}.
They were particularly interested in Type I migration rates on T Tauri discs, and as part of this, estimated the minimum gap opening planet in their models.
Our model is showing similar behaviour, with the minimum mass required for gap opening relatively flat out to $\sim 10 \, \textrm{au}$, and rising thereafter.
This is despite the numerous differences in the disc and dust models, and the gap opening criteria used.
\citet{2005ApJ...619.1123J} also noted that the minimum planet mass required to open a gap would be lower nearer to the star.

Because the minimum gap opening planet mass ratio is less sensitive to radius in the inner regions, the transition radius ($r_{tr}$, Equation~\ref{eqn:rtr}), between viscous heating and heating by stellar radiation sets a favourable spot for a planet migrating inward via type I migration to slow its migration rate.
Type II migration, following gap opening, is expected to be slower than type I migration for Earth mass objects, lacking a gap \citep{2000prpl.conf.1135W}.
A planet migrating in the outer disk could become able to open a gap as it moved inward.
The planets that survive would perhaps be the ones that were just massive enough to open gaps in the disc at this transition radius; planets unable to open gaps would continue to migrate rapidly, and be lost onto the star.
If a flared disk does not contain a gap, then the minimum gap opening planet mass ratio computed at this transition radius, $r_{tr}$ (Equation~\ref{eqn:rtr}) provides an estimate for its minimum mass.

As the minimum gap opening planet mass is approximately equal to $q_0$,  we can compute how it scales with radius, stellar mass, luminosity, accretion rate and $\alpha$ parameter.
For heating from viscous dissipation and an optically thick disc, the minimum gap opening planet mass ratio is
\begin{equation}
\begin{split}
q_{min,\nu,1} &\sim 
     50 \parenfrac{\alpha k_B}{\mu m_H}^{\frac{4}{5}}
     \parenfrac{3 \kappa_V \dot{M}^2}{128 \pi^2 \sigma}^{\frac{1}{5}}
     \paren{GM_*}^{-\frac{7}{10}} 
     r^{\frac{1}{10}} \\
  & \sim
     \scinot{4}{-4}
     \parenfrac{\alpha}{0.01}^{\frac{4}{5}}
     \parenfrac{\dot{M}}{10^{-8} M_\odot/{\rm yr}}^{\frac{2}{5}} 
     \times \\
 &     \parenfrac{M_*}{M_\odot}^{-\frac{7}{10}}
     \parenfrac{r}{\rm AU}^{\frac{1}{10}}
\end{split}
\label{eqn:qminnut1}
\end{equation}
The minimum gap opening planet mass ratio is only weakly dependent on radius (with exponent $1/10$) and most strongly dependent on $\alpha$ and $M_*$.
For heating from viscous dissipation and an optically thin disc, the minimum gap opening planet mass ratio
\begin{equation}
\begin{split}
q_{min,\nu,0} & \sim
        50  \alpha \parenfrac{ k_B}{\mu m_H}
       \parenfrac{3 \dot{M} T_\odot}{8 \pi \sigma}^{\frac{1}{5}}
       \paren{GM_*}^{-\frac{4}{5}} 
       r^{\frac{2}{5}}  \\
  &\sim
     \scinot{4}{-4}
     \parenfrac{\alpha}{0.01}
     \parenfrac{\dot{M}}{10^{-8} M_\odot/{\rm yr}}^{\frac{1}{5}}
     \times \\
 &
     \parenfrac{M_*}{M_\odot}^{-\frac{4}{5}}
     \parenfrac{r}{\rm AU}^{\frac{2}{5}}
\end{split}
\label{eqn:qminnut0}
\end{equation}
The minimum gap opening planet mass ratio is more strongly dependent on radius  when the disk is optically thin than optically thick.

When heating is due to irradiation by the star, we find
\begin{equation}
\begin{split}
q_{min,i} & \sim
     50 \alpha
     \parenfrac{k_B}{\mu m_H}^{\frac{10}{9}}
     \parenfrac{L_*(1-\beta) T_{\odot}}{12 \pi \sigma}^{\frac{2}{9}}
     \times \\
     & 
     \paren{GM_*}^{-\frac{10}{9}} 
     r^{\frac{2}{3}} \\
  &\sim
     \scinot{6}{-4}
     \parenfrac{\alpha}{0.01}
     \parenfrac{L_*}{L_\odot}^{\frac{2}{9}}
     \parenfrac{M_*}{M_\odot}^{-\frac{10}{9}}
     \parenfrac{r}{\rm AU}^{\frac{2}{3}}
\end{split}
\end{equation}
The radial exponent here is 2/3, larger than for either the high and low opacity viscous heating cases.
For the irradiated disk $q_{min}$ depends upon the $\alpha$ parameter because the gap opening criterion depends on the disk viscosity, however $q_{min}$ does not depend on the accretion rate $\dot M$ because the disk temperature is independent of $\dot M$.

Were the disk to become self-shadowed then the solid lines
on Figure~\ref{fig:minimumplanetmdot} and subsequent figures are relevant.
The minimum gap opening planet mass does not significantly change, but the radius at which a low mass object can open a gap is further out and set instead by the location of $r_\tau$ (Equation~\ref{eqn:rtau1}).
If the disk is self-shadowed then the radius at which the disk becomes optically thin, $r_\tau$, might set a favorable spot for an inward migrating core to open a gap.

We compute the minimum gap opening planet mass ratio for an optically thick disc at the transition radius $r_{tr}$ finding 
\begin{multline}
q_{min}(r_{tr}) \sim
     \scinot{4}{-4}
     \parenfrac{\alpha}{0.01}^{0.8}
     \parenfrac{L_*}{L_\odot}^{-0.08} \times \\
     \parenfrac{\dot{M}}{10^{-8} M_\odot/{\rm yr}}^{0.48}
     \parenfrac{M_*}{M_\odot}^{-0.58}
 \label{eqn:qminrtr}
\end{multline}
This can be used to place a constraint on possible planets residing in a
gapless disk,
\begin{multline}
M_p \la
     0.4 \, M_J \,
     \parenfrac{\alpha}{0.01}^{0.8}
     \parenfrac{L_*}{L_\odot}^{-0.08} \times \\
     \parenfrac{\dot{M}}{10^{-8} M_\odot/{\rm yr}}^{0.48}
     \parenfrac{M_*}{M_\odot}^{0.42}
  \label{eqn:Mp}
\end{multline}
where $M_J$ is a Jupiter mass.
The above minimum gap opening planet mass ratio and mass are
appropriate for both flared and self-shadowed disks because
the minimum gap opening planet mass is insensitive
to radius within the transition radius.

\begin{figure}
\includegraphics[angle=0,width=3.0in]{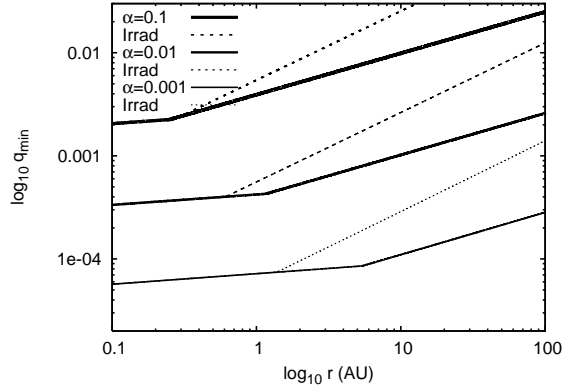}
\caption{\label{fig:minimumplanetalpha} 
Similar to Figure~\ref{fig:minimumplanetmdot} except the minimum
gap-opening planet
mass is shown as a function of radius for a disk with 
$\alpha=0.001, 0.01, 0.1$, with 
accretion rate  $\dot{M}=10^{-8} M_{\odot}/$yr and stellar mass 
$M_* =1 M_{\odot}$.  
The solid lines are for the case of viscous heating only 
and the dotted lines are for the case of irradiated heating only.
}
\end{figure}

In Figure~\ref{fig:minimumplanetalpha} we show the minimum
gap-opening planet mass ratio
for three disks with the same accretion rate, $\dot M = 10^{-8}M_\odot$/yr,
and stellar mass, $M_* = 1 M_\odot$.
but for three different $\alpha$ parameters, $\alpha = 0.1, 0.01$ and 0.001.
Figure~\ref{fig:minimumplanetalpha} can be directly compared
to Figure~\ref{fig:minimumplanetmdot} as the central solid 
and dotted lines are the same.
We see that lower mass planets can open a gap in 
lower $\alpha$ disks. We note that the minimum gap opening
planet mass is more strongly sensitive to this poorly constrained
parameter, $\alpha$ than the other parameters, such as stellar
mass, luminosity and accretion rate.

\begin{figure}
\includegraphics[angle=0,width=3.0in]{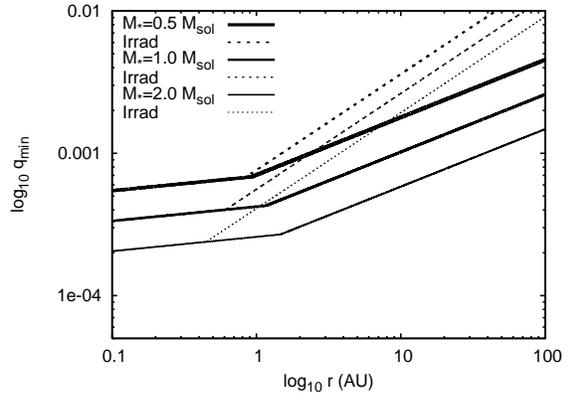}
\caption{\label{fig:minimumplanetstellarmass} 
Similar to Figure~\ref{fig:minimumplanetmdot}
and~\ref{fig:minimumplanetalpha} 
except the minimum gap-opening planet ratio for different
stellar masses
$M_* = 0.5, 1, 2$ $M_{\odot}$ for a disk
with $\alpha=0.01$ and accretion rate $\dot{M}=10^{-8} M_{\odot}/$yr.
The solid lines are for the case of viscous heating only 
and the dotted lines are for the case of irradiated heating only.
We have assumed that $L_* = L_\odot (M_*/M_\odot)^3$.
}
\end{figure}

In Figure~\ref{fig:minimumplanetstellarmass} we show the minimum gap-opening planet mass ratio for three disks with the same accretion rate, $\dot M = 10^{-8}M_\odot$/yr, and $\alpha=0.01$ but with different stellar masses, $M_* = 0.5, 1.0$ and $2.0 M_\odot$.  
Figure~\ref{fig:minimumplanetstellarmass} can be directly compared to Figures~\ref{fig:minimumplanetmdot} and~\ref{fig:minimumplanetalpha}.
The luminosity of a star depends on the stellar mass, and here we have assumed that $L_* = L_\odot (M_*/M_\odot)^3$ to take this into account.
Figure~\ref{fig:minimumplanetstellarmass} shows that the lower mass stars (rather than higher mass ones) require higher planet mass ratios to open a gap.

\begin{figure}
\includegraphics[angle=0,width=3.0in]{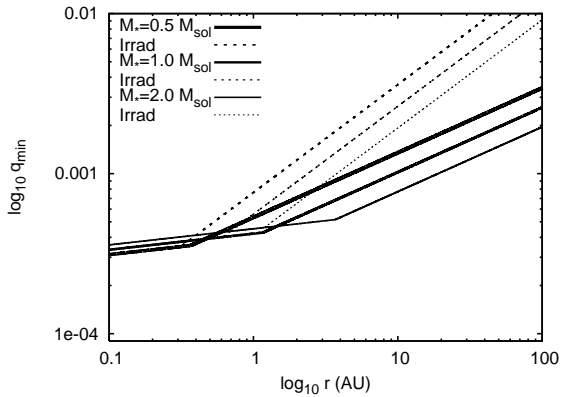}
\caption{\label{fig:minimumplanetstardisc} 
The minimum planet mass ratio for
stellar masses $M_* = 0.5, 1, 2$ $M_{\odot}$ for disks with $\alpha=0.01$.
We have assumed that $\dot{M}=10^{-8}~M_\odot/{\rm yr}~(M_*/M_\odot)^2$
and $L_* = L_\odot (M_*/M_\odot)^3$.
The solid lines are for the case of viscous heating only 
and the dotted lines are for the case of irradiated heating only.
}
\end{figure}

Figure~\ref{fig:minimumplanetstardisc} shows that the minimum planet mass ratio  for three different mass stars with the additional requirement that the accretion rate  is proportional to the stellar mass $\dot M \propto M_*^2$,  as suggested by observational surveys \citep{2005ApJ...625..906M}.
With this assumption we find that the minimum gap opening planet mass ratio is approximately independent of stellar mass.
Assuming that stellar luminosity scales with mass, $L_* = L_\odot (M/M_*)^3$  and $\dot{M} = 10^{-8} M_\odot/{\rm yr} \parenfrac{M_*}{M_\odot}^2$, Equation~\ref{eqn:Mp} becomes
\begin{equation}
M_p \la   0.4 M_J 
     \parenfrac{\alpha}{0.01}^{0.8}
     \parenfrac{M_*}{M_\odot}^{1.14}
\end{equation}
leaving only $\alpha$ as an undetermined parameter.
If larger planets are formed earlier during epochs of higher accretion around more massive stars then we would predict that they might be more likely to be located at larger radii.
This follows because the transition radius is larger for higher mass stars (Equation~\ref{eqn:rtr}) as is $r_{\tau}$ (Equation~\ref{eqn:rtau1}).

The favorable location for an inward migrating planet to
migrate inward for a flared disk assuming
$L_* = L_\odot (M/M_*)^3$
and $\dot{M} = 10^{-8} M_\odot/{\rm yr} \parenfrac{M_*}{M_\odot}^2$
would be  the transition radius computed for these assumptions
\begin{equation}
   r_{tr} \sim 0.3 {\rm AU} \parenfrac{M_*}{M_\odot }^{0.17}
\label{eqn:rtracc}
\end{equation}
(modifying Equation~\ref{eqn:rtr}) and that for a self-shadowed disk 
\begin{equation}
r_{\tau} \sim 1.2 {\rm AU} \parenfrac{\alpha}{0.01}^{\frac{2}{3}}
           \parenfrac{M_*}{M_\odot}^{\frac{5}{3}}
\end{equation}
(modifying Equation~\ref{eqn:rtau1})
Though the minimum gap opening planet mass is not significantly different in the self-shadowed disk than the flared one, lower mass planets can open gaps at larger radius.

If larger planets are formed earlier during epochs of higher accretion around more massive stars then we would predict that they might be more likely to be located at larger radii.
This follows because the transition radius is larger for higher mass stars.

\section{Discussion and Summary}
\label{sec:discuss}

In this paper we have explored the relation between the
minimum gap-opening planet mass and disk parameters,
accretion rate, the alpha parameter and stellar mass.
We have used simple models for $\alpha$ discs, with temperature set by heating due to viscous accretion and stellar irradiation.

We find that the minimum gap opening planet mass is not strongly sensitive to radius at radii where the disk is heated primarily by viscous dissipation and is optically thick
However, the minimum gap opening planet mass increases with radius where the disk is heated primarily by stellar radiation.
Due to the weak dependence with radius in an optically thick disk heated by viscous dissipation, the minimum mass gap opening planet in any disk is best estimated by the viscous criterion (Equation~\ref{eqn:qminnut1}).
For a self-shadowed disk, the minimum gap opening planet
mass is similar to that for a flared disk, however the radius at which
the mass begins to increase more rapidly is at larger radius and
is approximately where the disc becomes optically thick.

We estimate that a $0.4$ Jupiter mass planet is required to open a gap in a flared disk  with $\dot M = 10^{-8} M_\odot/$yr and $\alpha = 0.01$ around a $1M_\odot$ star. 
Lower mass planets can open gaps in disks
with lower accretion rates and lower $\alpha$ 
parameters around a lower mass star.
We estimate that the minimum gap opening planet mass is
proportional to 
$\dot M^{0.48} \alpha^{0.8} M_*^{0.42} L_*^{-0.08}$ 
(Equation~\ref{eqn:Mp}).
This scaling relation can be used to
place limits on planets residing in gapless disks as a
function of accretion rate, $\alpha$ parameter, stellar mass and stellar
luminosity for both flared and self-shadowed disks.

It is interesting to speculate on scenarios for planet migration.
A planet that is not sufficiently massive to open
a gap but continues to accrete and migrate could
open a gap if it accretes mass sufficiently rapidly
\citep{2007ApJ...656L..25T}
or because it reaches a location in the disk where
migration ceases \citep{2006ApJ...642..478M,2007MNRAS.377.1324C}.
Consider a planet which has formed far out in the disk, and is migrating inwards in the linear (Type I) regime.
If it reaches the transition radius ($r_{tr}$) before achieving the mass required for gap opening at that point (Equation~\ref{eqn:qminrtr}), then it will continue its rapid inward migration, and will probably be lost on to the star.
The weak dependence of the gap opening mass in the inner disk with radius means that it is unlikely that the planet will be able to accrete fast enough to open a gap within $r_{tr}$.
The transition radius represents a `last chance' for a planet to open a gap and slow its migration.
If a disk is observed without a gap, Equation~\ref{eqn:qminrtr} provides a bound on the most massive object in the inner portions of the disk.

The different models do predict different locations
for planets to slow their migration.   
In the planet trap proposed by \citet{2006ApJ...642..478M}
the truncation of the disk near the star sets the location
of the planet trap. As accretion would not be detected
after a planet formed, the scenario suggested
here would relate the mass of the first planet formed 
to the minimum detectable disk accretion rate. 
If planet formation cuts off disk accretion 
and planets stop migration at the transition radius then we would
predict a relation between clearing size and stellar mass.
Equation~\ref{eqn:rtracc} which relates
the transition radius to the stellar mass assuming
$L_* \propto M_*^2$ and $\dot M \propto M_*^3$
suggests that the radius of the clearing following 
the first gap opening planet would
grow with the stellar mass.
We note that the radii, $r_{tr}$ and $r_\tau$, estimated here are of
order 1 AU at an accretion rate of $10^{-8} M_\odot$/yr
and so smaller than hole clearing radii measured for objects such
as CoKuTau/4. 
The transition radius is only as large as CoKuTau/4's ($\sim 10$AU)
at an accretion rate of order $\dot{M} \sim 10^{-7}M_\odot$/yr.

The disk models considered here are simplistic
and do not include the radiative transfer of
more sophisticated models such as
\citet{2001ApJ...553..321D,2007ApJ...654..606G}
which could be used to improve upon the accuracy of
our constraints on the minimum gap-opening planet mass.
Better observational constraints on the $\alpha$ viscosity parameter
would also improve these estimates.
We have explored disks with opacity with different
temperature laws for gas disks using the opacities
described by \citet{1994ApJ...427..987B,1997ApJ...486..372B}
The shapes of
the curves differ quantitatively but not qualitatively from
those shown here.
In each temperature regime the disk temperature has
a different slope but the overall shape of the temperature, aspect
ratio and density curves are similar to and within
a factor of a few of those shown here.
We have not shown these curves here as the dust opacity is expected to dominate that of the gas, and the simple model explored here can be solved more easily to produce our scaling relations.

Here we have only considered models of steady state disks with constant accretion rate $\dot M$.
A starved disk that has not enough mass at large radius to maintain its accretion rate would have lower disk density at larger radius than predicted here.
If the disk were flared, the disk temperature would be set by stellar irradiation and the minimum gap-opening planet mass estimate would not vary from what is predicted here.
However were the disk to become self-shadowed the minimum gap opening planet mass would be lower at than that predicted with a constant $\dot{M}$ (this follows because $q_{min}$ decreases with decreasing $\dot M$; Equations~\ref{eqn:qminnut1} and~\ref{eqn:qminnut0}).

Assuming that the accretion processes within the disc can be fully parameterised by $\alpha$ is questionable.
The source of accretion disc viscosity is poorly understood, but the most likely candidate is the magneto-rotational instability (MRI).
\citet{2003ApJ...589..543W} performed magneto-hydrodynamic (MHD) calculations of a planet embedded in a disc, and found that the gap structure was markedly different to that obtained from a calculation which included a physical viscosity.
We have also neglected the possibility of a `dead zone' in the disc \citep{1996ApJ...457..355G}, where the high midplane densities shut down the MRI.
The surface layers continue to be ionised by cosmic rays, and can still accrete.
Within the dead zone, the effective viscosity is expected to be dramatically lower, and hence much lower mass planets might be able to open gaps.
In this case, the first (tidal) term of Equation~\ref{gc1} would become more significant in determining whether a planet can open a gap.
This is in contrast to the results we have presented here, which are dominated by the second (viscous) term.
\citet{2005ApJ...618L.137M,2006MNRAS.365..572M} considered gap opening in a disc with a dead zone.
Figure~2 of the \citeyear{2005ApJ...618L.137M} paper shows their expected minimum mass gap opening planets, as a function of radius.
\citeauthor{2005ApJ...618L.137M} found that the minimum mass planet required to open a gap was generally increasing with radius (unlike the nearly constant value in the inner regions we calculated above), with a step at the boundary of the dead zone.
There are numerous differences between their work and ours; not only are the disc models different, but also the gap opening criterion.
A criterion calculated by \citet{2002ApJ...572..566R} is used, which is based on a calculation of spiral shock dissipation in circumstellar discs.
The structures formed by a planet which satisfies the viscous, but not the tidal, condition for gap opening (that is, the first term of Equation~\ref{gc1} is greater than 1, but the second term is much less than 1) have not been extensively studied numerically, and we intend to investigate them further in future work.
It should be noted that `dead zones' might not be dead as first thought.
Although the MRI cannot operate (due to the low ionisation levels), turbulence can diffuse downwards from the surface layers \citep{2003ApJ...585..908F,2007astro.ph..2549O}, stirring material and driving accretion.

In this paper, we have calculated the minimum mass planet capable of opening a gap in a steady state accretion disc.
In the inner regions, the minimum mass required is very weakly dependent on radius.
Further out, the minimum mass grows with radius.
The transition between the two regimes (located where heating by stellar radiation dominates, or where a self-shadowed disc becomes optically thin) represents a `last chance' for a rapidly migrating embedded planet to open a gap.
If a planet is unable to do this, it is likely to be lost onto the star.


\section*{Acknowledgements}

We thank A. Crida, A. Morbidelli, A. Frank and E. G. Blackman
for valuable and helpful discussions.
Support for this work was in part
provided by National Science Foundation grant AST-0406823,
and the National Aeronautics and Space Administration
under Grant No.~NNG04GM12G issued through
the Origins of Solar Systems Program,
and HST-AR-10972 to the Space Telescope Science Institute.


\bibliography{general}
\bibliographystyle{astron}

\bsp

\label{lastpage}

\end{document}